\documentclass{aa}
\usepackage{longtable,graphics,latexsym,amssymb,times,psfig,amsmath}

\def\ltsima{$\; \buildrel < \over \sim \;$}
\def\lsim{\lower.5ex\hbox{\ltsima}}
\def\gtsima{$\; \buildrel > \over \sim \;$}
\def\gsim{\lower.5ex\hbox{\gtsima}}

\def\sw{{\it Swift}}
\def\ba{BATSE}
\def\he{Hete--II}
\def\ep{$E_{\rm peak}$}
\def\epo{$E^{\rm obs}_{\rm peak}$}
\def\flu{$F$}
\def\pf{$P$}
\def\liso{$L_{\rm iso}$}
\def\eiso{$E_{\rm iso}$}
\def\ama{$E_{\rm peak}-E_{\rm iso}$}
\def\yone{$E_{\rm peak}-L_{\rm iso}$}

\begin{document}

\title{Short versus Long Gamma--Ray Bursts: spectra, energetics, and luminosities}

\author{G. Ghirlanda\inst{1}
\and L. Nava\inst{1,2}
\and G. Ghisellini\inst{1}
\and A. Celotti \inst{3}
\and C. Firmani \inst{1,4}}

\offprints{Giancarlo Ghirlanda \email{giancarlo.ghirlanda@brera.inaf.it}}

\institute{Osservatorio Astronomico di Brera, via E. Bianchi 46, I--23807
  Merate (LC), Italy 
\and Universit\`a degli Studi dell'Insubria, Dipartimento  di Fisica e Matematica, 
  via Valleggio 11, I--22100 Como, Italy 
\and SISSA, Via Beirut 2/4, I--34014, Trieste, Italy 
\and Instituto de Astronomia, Universidad Nacional Autonoma de Mexico, A.P. 70-264, 04510 Mexico D.F., Mexico}
   
\date{Received ... / Accepted ...}  

\titlerunning{Short versus Long GRBs} 
\authorrunning{Ghirlanda et al.}

\abstract{We compare the spectral properties of 79 short and 79 long
  Gamma--Ray Bursts (GRBs) detected by BATSE and selected with the same
  limiting peak flux.  Short GRBs have a low--energy spectral component harder
  and a peak energy slightly higher than long GRBs, but no difference is found
  when comparing short GRB spectra with those of the first 1--2 sec emission
  of long GRBs. These results confirm earlier findings for brighter GRBs.  The
  bolometric peak flux of short GRBs correlates with their peak energy in a
  similar way to long bursts.  Short and long GRBs populate different
  regions of the bolometric fluence--peak energy plane, short bursts being
  less energetic by a factor similar to the ratio of their durations. If
  short and long GRBs had similar redshift distributions, they would have
  similar luminosities yet different energies, which correlate with the peak
  energy \ep\ for the population of long GRBs.  We also test whether short
  GRBs are consistent with the \ama\ and \yone\ correlations for the available
  sample of short (6 events) and long (92 events) GRBs with measured redshifts
  and \epo: while short GRBs are inconsistent with the \ama\ correlation of
  long GRBs, they could follow the \yone\ correlation of long bursts.  All the
  above indications point to short GRBs being similar to the first phases of
  long bursts. This suggests that a similar central engine (except for its
  duration) operates in GRBs of different durations.  
} 
\maketitle

\section{Introduction}

Since the launch of the \sw\ satellite (Gehrels et al. 2004), several 
pieces of information have been added to the puzzle concerning short GRBs (e.g. see Nakar
2007 and Li \& Ramirez--Ruiz 2007 for recent reviews). Short GRBs exhibit X--ray and
optical afterglows, similar to those of long GRBs, and in a few cases also X--ray
flares, similar to those discovered in the class of long events (e.g. Barthelmy
et al. 2005). Short bursts have, on average, lower fluences and similar peak
fluxes of long GRBs. Their X--ray and optical afterglows scale with the fluence
(Gehrels et al. 2008; Nysewander, Fruchter \& Peer, 2008). The redshift
distribution of short GRBs is still an open issue due to the few secure $z$
measured to date. Statistical studies (e.g.  Magliocchetti, Ghirlanda \&
Celotti, 2003, Tanvir et al. 2005, and Ghirlanda et al. 2006) seem to imply that a
significant fraction of the \ba\ short bursts are located in the local
universe, while direct $z$ measurements, in the \sw-era, suggest an average
$\langle z \rangle\sim 1.0$.  The discovery of the intrinsically short
($T_{90}^{\rm rest}\sim$ 1s) GRB~080913 is even more challenging, being the
most distant GRB to date ($z=6.7$, Fynbo et al.  2008).

Understanding the nature of the host galaxies of short GRBs is also a 
challenge (e.g. Berger 2006): if they originate from a merger of two compact (evolved) objects, they should be
preferentially located in early--type galaxies (although see Belczynski et al.
2008).  \sw\ observations appear to infer that the formal separation at about 2 seconds
in the (observed) duration distribution of short and long bursts might not be
correct.  It was known from \ba\ and \he\ (Norris \& Bonnell 2006;
Donaghy et al. 2006) that the short hard spikes can be followed by dim,
very long--duration emission (referred to as ``extended emission'').
Short--spikes with extended emission were also found in the population of
\sw\ GRBs (e.g. Norris \& Gehrels 2008).  It remains unclear whether these
events represent a third category in the temporal classification of a different
origin (e.g. see Zhang et al.  2007; Della Valle et al. 2006 for the case of
GRB 060614).

Short GRBs have been assumed to differ from long events on the
basis of their different properties in the hardness ratio--duration plot
(Kouveliotou et al. 1993). However, the hardness ratio is only approximately
representative of the burst spectral properties.  By completing a detailed analysis
of the spectra of bright \ba\ short bursts, Ghirlanda, Ghisellini \& Celotti
2004 (GGC04 hereafter) showed that their spectra are harder than those of long
GRBs, due to a harder low--energy spectral component, rather than a different
peak energy.  GGC04 also found that the spectra corresponding to the first 1--2 seconds of
emission of long GRBs are similar to those of short bursts. This result
relies on the detailed spectral modelling of GRB spectra rather than on the
hardness--ratio analysis (but see Dong \& Quin 2005 and Quin \& Dong 2005).
Interestingly, also in the temporal domain, the properties of short GRBs appear similar to those during the
first seconds of the emission of long events: Nakar \& Piran (2002) found that
the typical variability timescale of short GRBs ($\sim$10 ms) corresponds to
that of the first 1--2 seconds of long ones.  These results might suggest a common
origin for the prompt emission of short and long GRBs.

The short burst sample analysed in GGC04 consists of the brightest 28 short GRBs
detected by \ba.  It is worth exploring whether the results hold when the spectral
analysis is extended to a significant number of short bursts with lower peak
fluxes.  In this respect, we note that for long \ba\ GRBs, the spectral
properties (such as \epo, i.e., the peak energy of the $\nu F_{\nu}$ spectrum)
correlate with their fluence and peak flux (Ghirlanda et al. 2008; Nava et al.
2008).

For the population of long GRBs with measured redshifts, the peak energy of the
prompt emission spectrum appears to correlate with the isotropic equivalent
energy \ama\ (so--called ``Amati'' correlation, from Amati et al. 2002) and/or with
the isotropic equivalent luminosity \yone\ (so--called ``Yonetoku''
correlation, from Yonetoku et al. 2004).

The interpretation of these correlations may provide additional insight into the nature
of the prompt emission. The few short GRBs with measured $z$ and well
determined spectral properties are inconsistent with the \ama\ correlation
(Amati 2006, 2008), but it is worth exploring whether they are consistent with
the \yone\ relation. It has also been shown that the rest--frame
correlations (\ama\ and \yone) for long GRBs correspond to observer--frame
correlations between the peak energy \epo\ and the fluence or peak flux (Nava
et al.  2008, N08 hereafter). Therefore, there are two possible tests that
can be performed: (a) compare short and long GRBs with respect to the observer
frame \epo-\flu\ and \epo-\pf\ trends; (b) compare (the still few) short and
long GRBs in the rest frame, where long GRBs define the \ama\ and \yone\ 
correlations.

The paper is organised as follows: in Sect. \ref{analysis}, we present the
results of the spectral analysis of a sample of short and similarly
selected/analysed long \ba\ GRBs with peak flux $>3$ phot cm$^{-2}$ s$^{-1}$;
in Sect. \ref{comparison}, the short GRB and long \ba\ GRB spectra are
compared; in Sect. \ref{correl}, we study the spectral--energy correlations for
the population of short and long GRBs and in Sect. \ref{restframe}, short GRBs
with known redshifts are compared with the \ama\ and \yone\ correlations
defined by the most updated sample of long GRBs. We discuss our findings in
Sect. \ref{concl}.

\section{Sample selection and spectral analysis}
\label{analysis}

\subsection{Short GRBs}

GGC04 considered 28 short \ba\ GRBs with peak flux exceeding $P>$ 10 phot
cm$^{-2}$ s$^{-1}$ (in the 50-300 keV energy range). To extend this
analysis, we selected a sample of short duration GRBs ($T_{90\%}< 2$ s) from
the \ba\ on--line catalogue\footnote{http://cossc.gsfc.nasa.gov/cossc/batse/}
with peak flux $>$ 3 phot cm$^{-2}$ s$^{-1}$ (integrated in the energy range
50-300 keV, and computed on a 64 ms timescale). In the sample of 497
triggered short \ba\ events with tabulated duration, peak flux and fluence
(see Magliocchetti et al. 2003), 157 short bursts satisfy the above selection
criterion.  This sample, used in Lazzati, Ghirlanda \& Ghisellini (2005),
also includes the 28 short \ba\ bursts studied by GGC04.

We analysed the Large Area Detector (LAD) spectral data of these GRBs.  For
13/157 GRBs either we could not find the data (6 cases) or the data were affected
by gaps (7 cases).  Seventy-nine of the remaining 144 GRBs have data with a sufficient
signal-to-noise (S/N) to fit the spectrum and constrain the spectral
parameters, while the low S/N ratio spectrum after background subtraction for
the other 65 short bursts does not allow a meaningful spectral fitting. For
these 65 cases, we attempted rebinning the spectra at $>$ 1$\sigma$  to
increase the signal, albeit at the expense of the spectral
resolution\footnote{The LAD spectra typically consist of  80--100 usable
  channels distributed in the energy range $\sim$30 keV--1.5 MeV.}.
However, for 46/65 events, the spectrum contained only one or two points after
 rebinning, and for 19/65, to only 5 points, not allowing us to constrain the
spectral parameters of any fitted model.

\begin{table}
\caption{Average properties of the selected short and long GRBs.}  
\label{average} 
\centering
\begin{tabular}{c|ccc}
\hline
\noalign{\smallskip}
            &$\langle T_{90}\rangle$ &$\langle P_{50-300} \rangle$ &$\langle F \rangle$\\
            &s                       &phot cm$^{-2}$ sec$^{-1}$    &erg cm$^{-2}$ \\
\noalign{\smallskip}
\hline   
\noalign{\smallskip}
79 Long GRBs      &17      &11.5      &$1.7\times 10^{-5}$                             \\
79 Short GRBs     &0.73    &11.12     &$3.6\times 10^{-6}$          \\
\noalign{\smallskip}
\hline
\end{tabular}
\end{table}
The average duration, peak flux (integrated over the 50--300 keV
energy range), and fluence (for energies $>$25 keV) of the short and
long \ba\ GRBs are reported in Table \ref{average}.

The time--integrated spectrum was fitted with three spectral models typically
used to analyse \ba\ GRB spectra  (e.g. Kaneko et al. 2006): the Band model
(Band et al. 1993) consists of two smoothly joined power--law, the
cutoff power--law (CPL) and single power--law model (PL). Both the CPL and the
Band model exhibit a peak in their $\nu F_{\nu}$ spectrum if $\alpha>-2$ and
$\beta<-2<\alpha$, respectively, where $\alpha$ is the low--energy power--law
photon spectral index of the CPL and Band model and $\beta$ is the index of
the high--energy spectral component of the latter model.

In 71/79 events, the CPL model represented a good fit to the spectra.  For the
remaining 8 cases, no significant curvature was found within the observed energy
range and the best--fit function was given by the PL model: for these cases, only a lower
or upper limit to \epo\, depending on the value of the power--law index, could
be set and a lower limit to bolometric fluence and peak flux (integrating 
over only approximately the 20--1000 keV range) could be estimated.
In most cases, we could not constrain the high--energy power--law
spectral index of the Band model since the typically lower flux of the
high energy channels does not allow us to discriminate between a
power--law component or an exponential cutoff, in most fits the CPL
model was statistically more robust than the Band one simply because it has
one parameter less. As already discussed in GGC04, we could not
perform a time--resolved spectral analysis with the LAD data of short
\ba\ bursts: if and how their spectrum evolve during the burst
duration remains an open issue.

The results of the spectral analysis of the time integrated--spectra
of the 79 short GRBs are reported in Table \ref{tabshort} ordered for
decreasing peak flux (Col. 3, as reported in the \ba\ on--line
catalogue). The spectral parameters of the best--fit function (CPL or
PL model) are listed in Cols. 4--6.  In the final two columns, we
report the bolometric (1--10000 keV) fluence and flux estimated from
the best--fit model parameters.  When only a lower or upper limit on
the energy $E_0$ could be determined the fluences and peak energy
fluxes are computed in the observed energy range 20--1000 keV.

\subsection{Long GRBs}

To compare the spectral properties of short and long GRBs, we
considered long \ba\ bursts selected with the same criterion as for
the short ones.  We note that the available samples of \ba\ bursts with
spectral information (e.g. Preece et al. 2000, and Kaneko et al. 2006)
were selected with different criteria: in particular, an analysis by Kaneko 
et al. (2006) examined bursts selected according to either peak
flux or fluence as part of the aim of performing time--resolved spectral
analysis with a minimum number of spectra distributed within each
burst.

For this reason, we randomly extracted from 
the data set of 400 long \ba\ GRBs with peak flux $>$3 ph cm$^{-2}$ s$^{-1}$ a
representative sub--sample of 79 GRBs by requiring that they followed
the Log$N$-Log$P$ defined by the entire sample of long \ba\ bursts.

In this case we also excluded events with S/N ratio insufficient to
constrain properly the spectral parameters.  We performed both the
time--integrated and time--resolved spectral analysis, adopting the
models defined in Sect. \ref{analysis}.  In 34/79 cases, the time--integrated 
spectrum is well described by a CPL spectral shape.
However, for 44/79 GRBs the spectrum shows a high energy power--law
tail and the Band model provides a more reliable fit than the CPL
one.  Only in one case is the best fit model function a simple power--law.

The time--integrated spectral results of the 79 long GRBs are reported
in Table \ref{tablong}. The bolometric fluence (Col. 8) and peak flux
(Col. 9) are both estimated by integrating the best--fit model in the
1 keV--10 MeV energy range.  For one GRBs (trigger 6400) the spectrum
is well fitted by a simple power--law. In this case we estimated only
a lower limit on fluence and peak flux.

\section{Short versus Long: spectra}
\label{comparison}

\subsection{Time integrated spectra}

The distribution of short and long \ba\ bursts in the hardness
ratio--duration plane (e.g. Kouveliotou et al. 1993) implies that
long and short GRBs are two separate classes with short GRBs being on
average harder than long ones.  However, similarly to long GRBs, the
spectrum of short GRBs presents a significant curvature in the \ba\
spectral range.
GGC04 compared the low--energy power--law index $\alpha$ and
the peak energy \epo\ of 28 bright ($P \ge$ 10 ph
cm$^{-2}$ s$^{-1}$) short bursts with those of long bursts selected on
the basis of a similar limiting peak--flux (Ghirlanda et al., 2002).
Despite the small sample, their results suggest that a statistically
significant difference existed in the low--energy part of the spectrum, such that 
short bursts had a harder spectral index $\alpha$: the average values
being $\langle \alpha_{\rm short} \rangle=-0.58\pm 0.10$ and $\langle
\alpha_{\rm long} \rangle=-1.05\pm 0.14$ (K--S probability $P_{\rm
KS}\sim$0.04\%).  The peak energy \epo\ of long bursts
($\langle E^{\rm obs}_{\rm peak} \rangle=520\pm 90 \rm ~keV$) is only
slightly higher than that of short events ($\langle E^{\rm obs}_{\rm
peak} \rangle=355\pm 30 \rm~keV$) with $P_{\rm KS}\sim$0.8\%.
\begin{table*}
\caption{Parameters of the Gaussian fits of the distributions of the spectral parameters of short and long GRBs.}
\label{tabhisto}
\centering
\begin{tabular}{ccccc|cccc}
\hline
\noalign{\smallskip}
   &     &$\alpha$&         &         &     &$E_{peak}$&          &         \\
   &short&long     &long 1 sec&long 2 sec&short&long      &long 1 sec&long 2 sec\\
\noalign{\smallskip}
\hline   
\noalign{\smallskip}
$\mu$    & -0.40 & -0.92 & -0.65 & -0.63 & 2.60  & 2.33  & 2.49  & 2.48  \\
$\sigma$ & 0.50  & 0.42  & 0.50  & 0.45  & 0.42  & 0.33  & 0.35  & 0.36  \\
\hline
\end{tabular}
\end{table*}

\begin{figure}
\hskip -0.5 true cm
\psfig{file=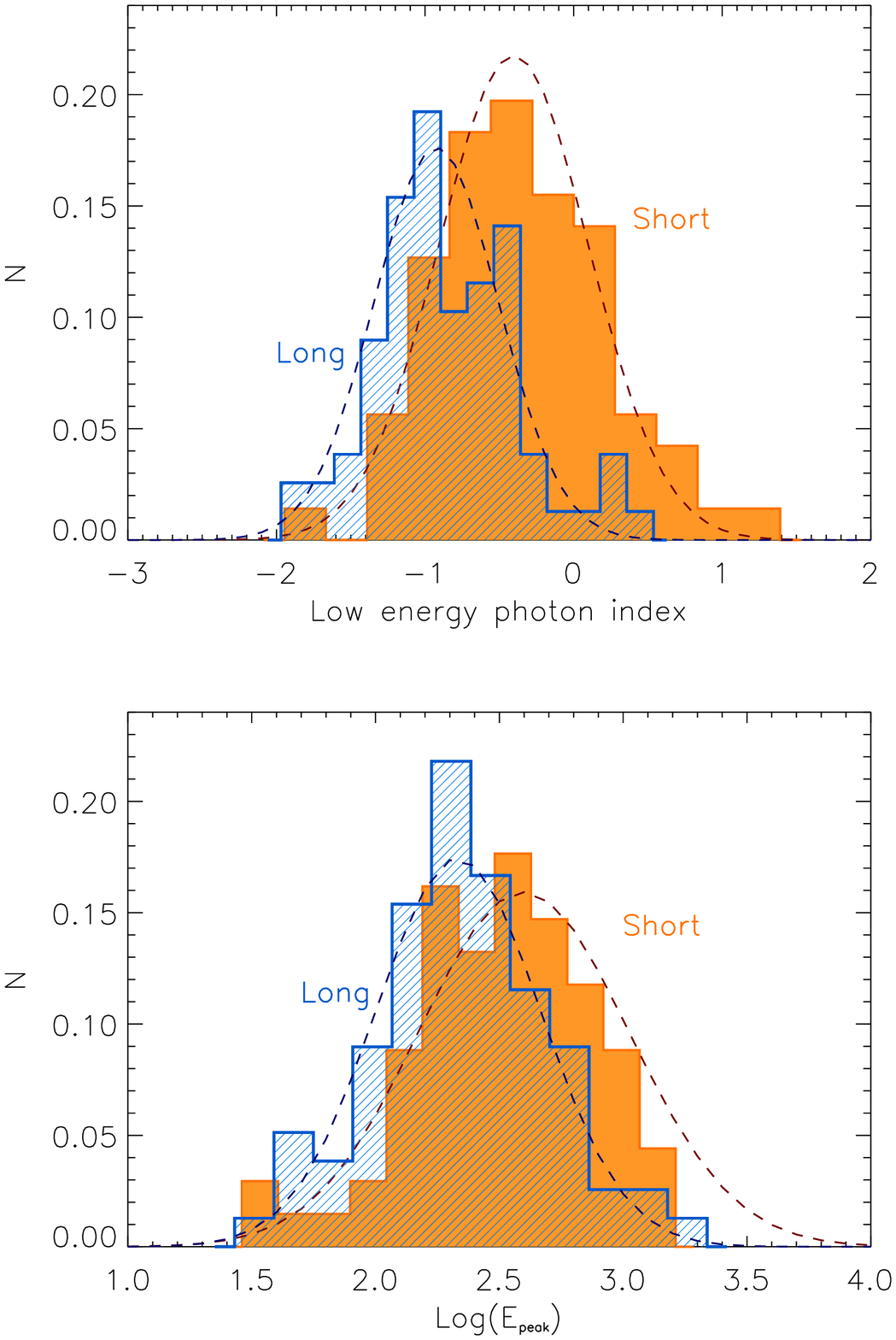,width=9.5cm,height=14cm}
\caption{Spectral parameter distributions normalised to the total
  number.  Top: low energy photon spectral index ($\alpha$) for the 79
  short (filled histogram) and the 79 long (hatched histogram) GRBs
  analysed in this work.  The dashed lines represent the Gaussian fit
  to these distributions.  Bottom: peak energy of the $\nu F_{\nu}$
  spectrum (\epo) for the two samples).
\label{histo}}
\end{figure}

We test these results using the larger samples of short and long GRBs
considered here, which extend the sample of GGC04 to the 3 ph
cm$^{-2}$ s$^{-1}$ peak flux threshold. The distributions of $\alpha$
and \epo\ for the time--integrated spectra of the 79 short and long
bursts are shown in Fig.  \ref{histo}.  They can be modelled well by
Gaussian functions: the best--fit model parameters ($\mu$ and
$\sigma$, representing the mean value and the standard deviation) are
reported in Table \ref{tabhisto}.  The K--S test probability that the
distributions were drawn from the same parent one are reported in
Table \ref{tabks}. For long events we report in Table \ref{tabks}
three cases: spectral parameters derived from the spectrum of
1) the whole emission, 2) the first second and 3) the first two
seconds of emission. For both $\alpha$ and \epo\ the probabilities
increase considering only the very first phases of long bursts.

The low--energy spectral index of short bursts is harder than that of
long GRBs $\langle \alpha_{\rm short} \rangle=-0.4\pm 0.5$ and
$\langle \alpha_{\rm long} \rangle=-0.92\pm 0.42$ ($P_{\rm
KS}=$8.8e-5), while their peak--energy
distributions are more similar ($P_{\rm KS}=$1.3\%). This result
confirms (see GGC04) that the spectral difference between short and
long GRBs as observed in the hardness--duration plane, is due to a
harder low--energy spectrum of short bursts rather than a
significantly different peak energy.

\subsection{Time resolved spectra of long bursts}

GGC04 found some evidence that the spectrum of short GRBs is similar
(in terms of $\alpha$ and \epo) to the spectrum of the first 1--2
seconds of the long events (the K--S probability of 83\% for $\alpha$ and
10\% for \epo). This result is intriguingly in agreement with the
findings that the variability timescale of short GRBs resembles that
of the first 1-2 seconds of long events (Nakar \& Piran 2002).

\begin{figure}
\hskip -0.5 true cm
\psfig{file=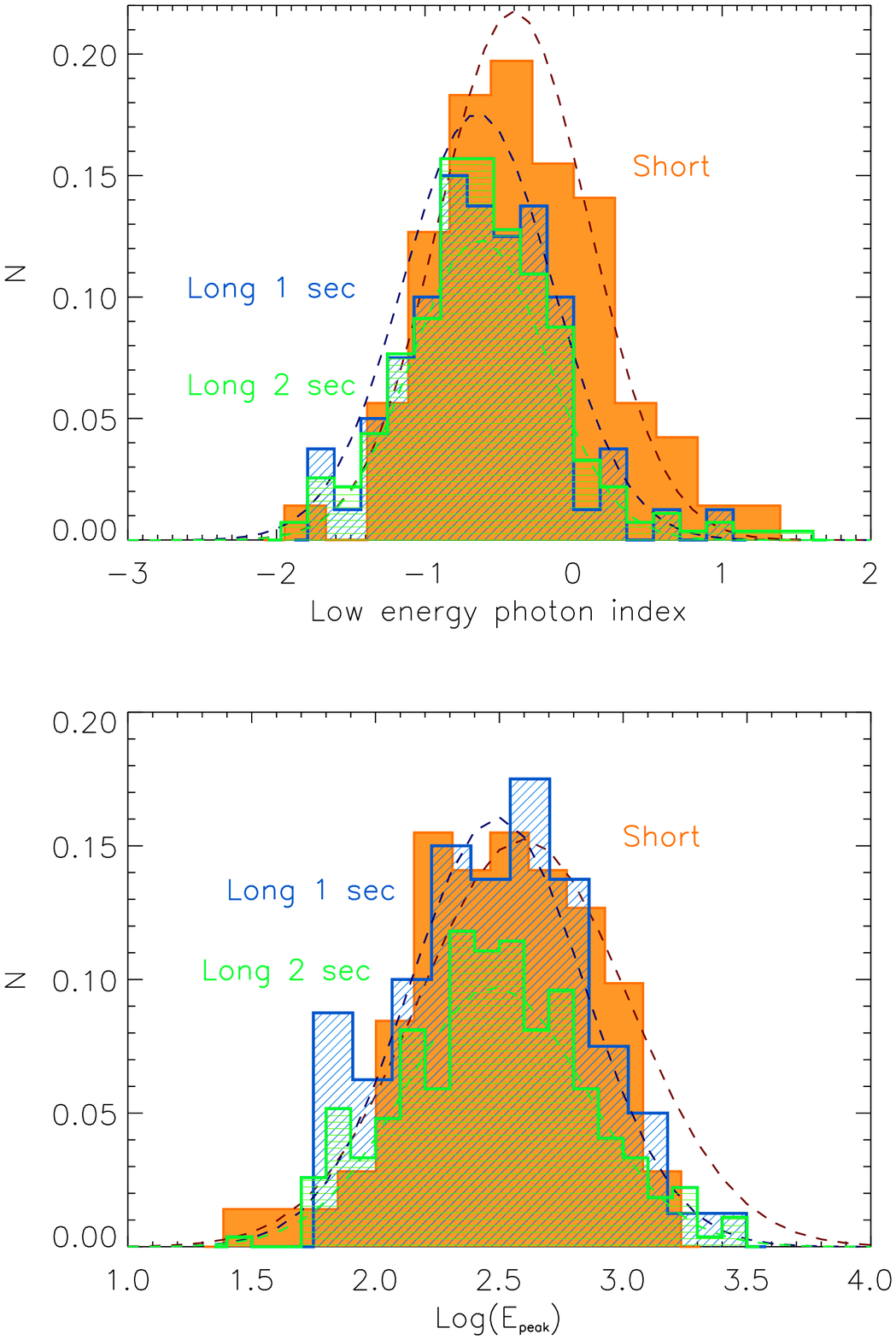,width=9.5cm,height=14cm}
\caption{Spectral parameter distributions normalised to the total
  number of spectra.  Top: low energy photon spectral index $\alpha$
  for the 79 short GRBs (filled histogram) and for the time resolved
  spectra of the first second (hatched oblique histogram) and the
  first two seconds (hatched horizontal histogram) of long bursts. The
  dashed lines represent the Gaussian fit to the distributions.
  Bottom: peak energy of the $\nu F_{\nu}$ spectrum for the same
  population of short and long bursts.
\label{histo12sec}}
\end{figure}

With the larger, uniformly analysed, samples of short and long GRBs
examined here, we can meaningfully compare the time--resolved spectra
of long GRBs during the first 2 secs of emission and the 
time--integrated spectra of short ones.  In Fig.  \ref{histo12sec}, we report
the $\alpha$ and \epo\ distributions of the 79 short bursts with those
determined by the time--resolved spectral analysis of long bursts
within one and two seconds of the trigger: the corresponding K--S
probabilities (see Table \ref{tabks}) indicate that they are very
similar.  The difference found between the time--integrated spectra of
short and long bursts could be due simply to a hard--to--soft
evolution of the spectrum of long GRBs, which become on average
softer with time than that of short bursts. While this should be tested by
comparing the spectral evolution of short and long GRBs, as
already found in GGC04 and mentioned above, the low S/N of the data used here prevents us
from performing a time--resolved spectral analysis for the short \ba\
bursts.

\section{Short versus Long GRBs: observer--frame correlations}
\label{correl}

Long bursts follow some empirical correlations involving the
(isotropic) energetics \eiso\ (Amati et al. 2002) and/or peak
luminosity \liso\ (Yonetoku et al. 2004), and the rest--frame peak
energy \ep.  N08 demonstrated that correlations also hold between
the observed peak energy \epo\ and the fluence (\flu) or observed peak
flux (\pf).  This opens the possibility of examining the impact of
instrumental selection effects on these correlations: in particular,
both the trigger threshold, i.e.  the minimum peak flux required to
trigger a given detector, and the ``spectral analysis'' threshold,
i.e. the minimum signal for a spectrum to be analysed, are functions
of \epo.  However, as N08 emphasized, the correlation between peak
energy and peak flux in the observer frame of \ba\ long GRBs is not
induced by these thresholds.

In comparing the distributions of the (observed)
\epo--fluence and \epo--peak flux for our representative sample of 79
short and long GRBs, we computed the bolometric (1--10000 keV) fluence
and peak flux by integrating the best--fit CPL or Band model.  When
the best fit model was instead a simple power--law, we can only
estimate a lower limit to the fluence (or peak flux) by integrating
the spectrum over the range 20--1000 keV and an upper/lower limit to
\epo, depending on the value of the fitted power--law index.

\begin{table} 
\centering
\caption{K--S test probability that the distributions of short and
  long GRB spectral parameters are drawn from the same parent
  population.}
\label{tabks}
\begin{tabular}{c|ccc}
\hline
\noalign{\smallskip}
                & long   &  long 1 sec  &  long 2 sec   \\
\noalign{\smallskip}
\hline   
\noalign{\smallskip}
$\alpha$ & 8.8e-7 & 0.0433     &  0.00457    \\
\ep\     & 0.013  & 0.965      &  0.764         \\
\hline
\end{tabular}
\end{table}

\subsection{\epo\ versus peak flux}

\begin{figure}
\psfig{file=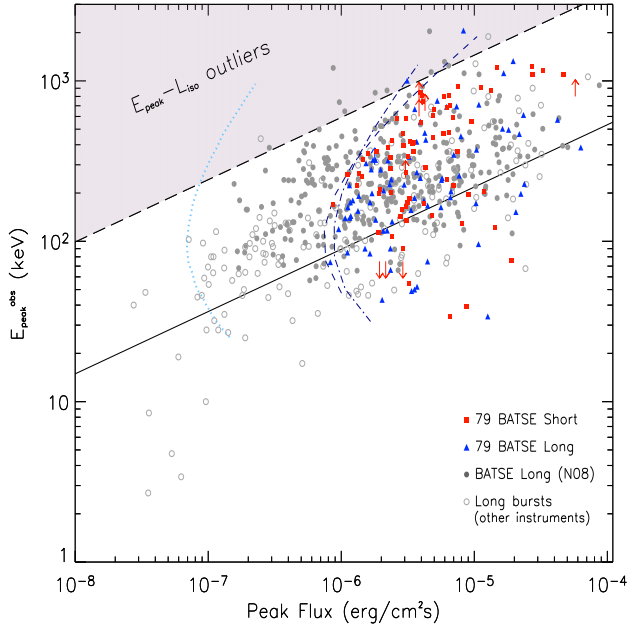,width=9cm,height=9cm}
\caption{Distribution in the \epo-\pf\ plane of the 79 long
  (triangles) and 79 short (squares) bursts considered here.  Arrows
  correspond to upper/lower limits to \epo, and in these cases the estimated
  \pf\ is a lower limit. For comparison also the \ba\ bursts from Kaneko et al.
  (2006) and N08 (filled circles) and the bursts detected by instruments other
  than \ba\ (empty circles -- see N08) are reported.  The dotted line
  represents the trigger threshold for \ba\ GRBs, i.e. the minimum peak flux
  needed to trigger the instrument.  The dot--dashed and dashed lines are the
  \pf\ limit criterion adopted to select the 79 long and 79 short GRBs,
  respectively: namely a photon peak flux of 3 phot cm$^{-2}$ s$^{-1}$ (in the
  energy range 50--300 keV) and a typical spectrum with $\alpha\simeq -0.5$
  and $\alpha\simeq -1.0$ for short and long GRBs, respectively. The solid
  line indicates the \yone\ correlation (as derived with the most updated
  sample of 92 GRBs with known $z$ in Sec.~\ref{restframe}) transformed in the
  observer frame \epo--\pf\ plane. The shaded region corresponds to a ``region
  of outliers'', namely values of \epo and \pf\ inconsistent (at more than
  3$\sigma$) with the \yone\ correlation for any GRB redshift.
  \label{pf-ep}}
\end{figure}

The 79 short and long GRBs populate similar regions of the \epo--\pf\
observer--frame plane\footnote{ \pf\ is the bolometric peak flux in erg cm$^{-2}$
  s$^{-1}$} (Fig. \ref{pf-ep}). For comparison in the figure,
we also report the complete sample of long GRBs (with fluence
$>2\times 10^{-6}$ erg cm$^{-2}$) analysed in N08 (filled circles) and
the (incomplete) samples of long bursts detected by instruments other
than \ba\ (open circles). The dotted line represents the trigger
threshold of \ba\ (adapted from Band 2006 -- see N08 for details).

To understand the impact of the selection threshold of the
sample (photon peak--flux $>$ 3 phot cm$^{-2}$ s$^{-1}$ between 50 and
300 keV) on the distribution of bursts in the \epo--\pf\
plane, we transformed the limiting photon--flux in bolometric
energy--flux by simulating different spectra with a variable \epo\ and
fixed typical values of $\alpha$ ($\alpha \simeq -1$ and $\simeq -0.5$
for long and short bursts, respectively), normalised to the photon
peak flux.  The corresponding curves (dot--dashed and dashed lines for
long and short GRBs, respectively) are shown in Fig.  \ref{pf-ep}: the
\ba\ trigger threshold is more than a factor of 10 lower than the
imposed selection criterion.
 
While the 79 long bursts confirm the existence of an \epo--\pf\
correlation independent of the instrumental effect due to the trigger
threshold (see N08), for short bursts the analysed range of peak flux
is insufficient to draw a definitive conclusion.  The selection cut
at low peak fluxes strongly affects the short--burst sample in the
\epo--\pf\ plane.  However, it is interesting to note that both
short and long GRBs -- selected with the same criterion -- are
consistent with the correlation defined by larger samples of long
events, which suggests the possibility that  short bursts also follow the
same (or a similar) \epo--\pf\ correlation exhibited by long
events.
    
In the \epo--\pf\ plane, we can also test the possible consistency of short
GRBs with the \yone\ correlation, namely $E_{\rm peak}\propto L_{\rm
  iso}^{0.4}$.  This correlation is updated and presented for short and long
GRBs in the next section, but  restricted to GRBs with known redshifts.
Assigning different redshifts to a GRB of unknown $z$, we define a trajectory
in the rest--frame \yone\ plane. This curve can intersect the
\yone\ correlation or become consistent with its 3$\sigma$ scatter. 
If not, the considered GRB is an outlier. 
Correspondingly, we can define the ``outlier'' region in the observer \epo--\pf\ plane 
 (see Nakar \& Piran 2005; Ghirlanda, Ghisellini \& Firmani
2005; N08).
This is the region where a GRB, regardless of  redshift, is inconsistent
with the \yone\ correlation, within its 3$\sigma$ scatter.
The shaded region in Fig.~\ref{pf-ep} represent this region: no short
burst of the 79 analysed is an outlier of the \yone\ correlation.

\begin{figure}
\psfig{file=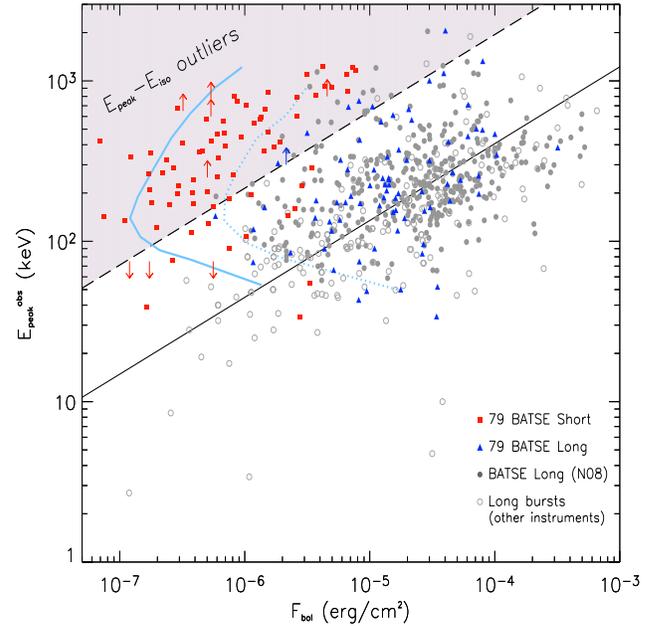,width=9cm,height=9cm}
\caption{Distribution of bursts in the \epo--\flu\ plane.  Squares
  (triangles) represent the sample of 79 short (long) bursts discussed
  in Sec.  \ref{analysis}.  The dotted and solid curves show the
  spectral threshold, i.e. the minimum fluence as a function of \epo
  necessary to perform a reliable spectral analysis and constrain the
  value of \epo itself. The threshold depends on the burst duration
  and on the spectral shape (see Ghirlanda et al. 2008 for more
  details): the dotted curve is estimated for long bursts while the
  solid curve is derived for a typical short burst of 0.7 sec duration
  and $\alpha=-0.5$ (as found in Sec.~\ref{analysis}). The solid line
  indicates the \ama\ correlation (as derived with the most updated
  sample of 92 GRBs with known $z$ in Sec.~\ref{restframe})
  transformed in the observer frame \epo--\flu\ plane. The shaded
  region represents the ``region of outliers'', namely values of \epo
  and \flu\ inconsistent (at more than 3$\sigma$) with the \ama\
  correlation for any redshift.
  \label{flu-ep}}
\end{figure}

\subsection{\epo\ versus fluence}

An analysis similar to that presented above can be performed by considering
the bolometric fluence. If the GRB redshift is known, from the bolometric
fluence one can derive the isotropic equivalent energy, which was found to be
correlated with the rest--frame peak energy (Amati et al. 2002, and Amati 2006).

In Fig.~\ref{flu-ep}, the 79 short and long \ba\ bursts (squares and
triangles, respectively) are reported together with the sample of long
\ba\ GRBs (filled circles) analysed by N08.  In contrast to our results for 
the \epo--\pf\ plane, short and long bursts occupy different
regions of the \epo--\flu\ plane, having similar \epo\ but fluences
scaling by a factor comparable to the ratio of their durations.  As
discussed by Ghirlanda et al. (2008), the observer frame \epo--\flu\
plane is  biased mostly by the ``spectral analysis'' threshold, which
corresponds to a requirement on the S/N in order to constrain the
spectral parameters.  The solid curve represents the ``spectral
threshold'' which is estimated for short bursts as described by Ghirlanda et
al. (2008): we adopted the typical value of $\alpha=-0.5$ determined in
Sect.~\ref{comparison} and the representative duration of the short
bursts included in our sample, namely $T_{90}\sim 0.7$ sec.  A burst
with \epo\ and \flu\ values so that it is located to the right of this curve has sufficient 
signal to allow a reliable spectral analysis.  The dotted curve in the figure
represents the ``spectral threshold'' for the population of long GRBs
(see N08 for the relevance of this selection effect to the properties
of long GRBs in the \epo--\flu\ plane).

Figure \ref{flu-ep} reveals that the spectral threshold affects the
distribution of short bursts significantly.  This was expected since
out of 144 bursts that satisfy the peak flux selection criterion and
with available data only for 79 the spectral parameters could be
constrained.  Due to these limitations, no conclusion can be inferred
about any true (i.e. not determined by selection effects) \epo--\flu\
correlation for short bursts.  However, it is clear from
Fig.~\ref{flu-ep} that short and long GRBs are highly scattered in the
\epo--\flu\ plane and that short GRBs do not follow the same
correlation defined by long events.

Finally, in the \epo--\flu\ plane, we can test the consistency of short
GRBs with the \ama\ correlation defined by long events (see
Sect.~\ref{restframe}). The region containing outliers is 
populated significantly: the majority of short GRBs ($\sim$78\%)
are inconsistent with the \ama\ correlation defined by long bursts.

\begin{figure*}
\psfig{file=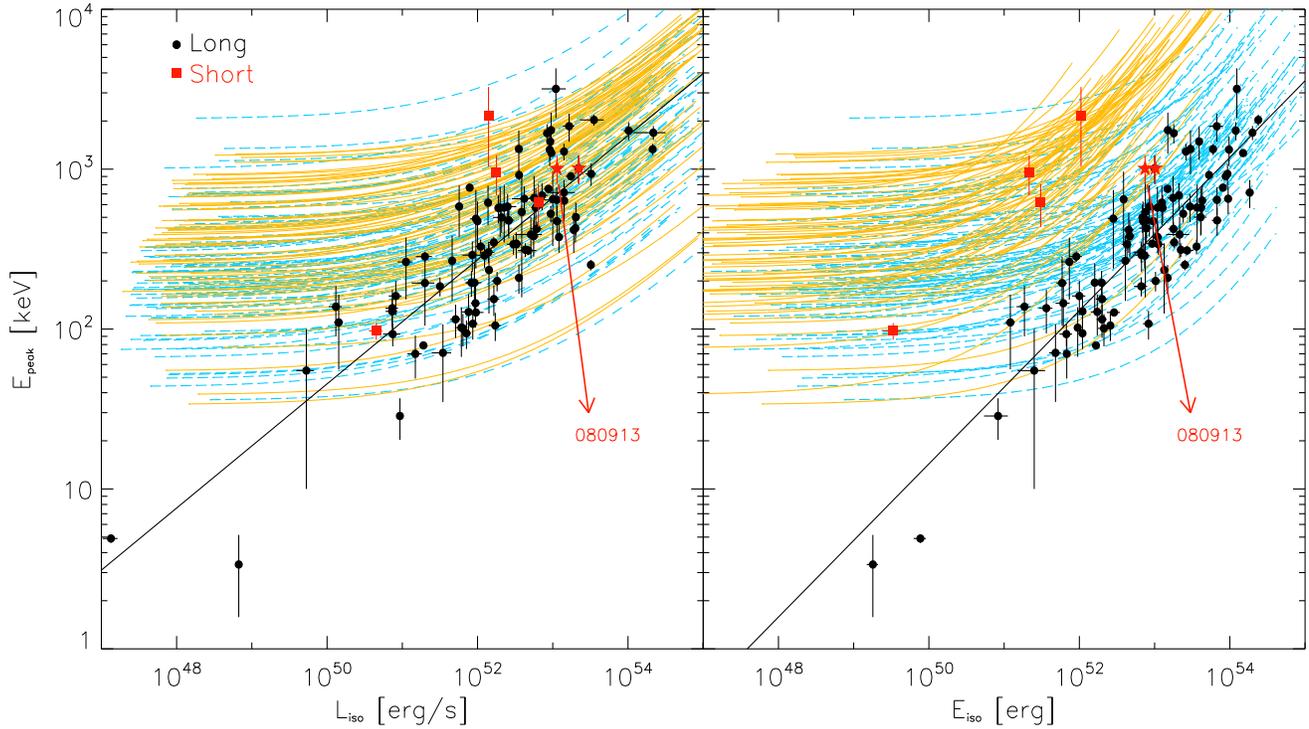,width=18cm,height=10cm}
\caption{Rest--frame peak energy versus isotropic luminosity (left) and
  isotropic energy (right) for 92 long (filled circles) and 6 short (filled
  squares and stars) GRBs with measured redshifts. The solid lines represent
  the best fit of the correlation defined by long GRBs: $E_{\rm peak}\propto
  L_{\rm iso}^{0.4}$ and $E_{\rm peak}\propto E_{\rm iso}^{0.5}$. The
  1$\sigma$ scatter is 0.23 dex and 0.27 dex for the \yone\ and \ama\ 
  correlation, respectively.  The orange and blue (dashed) lines indicate
  where the 79 short and long GRBs would be located for different redshifts
  (between 0.1 and 10). GRB~080913 (at $z$=6.7) and GRB~071020 (at $z$=2.145)
  are considered short events (filled stars) even though their observed
  duration is $\simeq$8 and $\simeq$4 sec, respectively.
  \label{ama_yone}}
\end{figure*}

\section{Short versus Long GRBs: energetics and luminosities}
\label{restframe}

The comparison between the data for short and long GRBs in the observer frame planes
has shown that although short and long bursts
have similar peak fluxes and peak energies, and can follow
the same correlation in the \epo--\pf\ plane, the distributions of
long and short GRBs are inconsistent in the \epo--\flu\ plane because
of the lower fluence of short GRBs.

In this section, we examine the isotropic energy and luminosity of the two populations 
and  consider in particular the two correlations, i.e. \ama\ and \yone, with similar
slopes and different normalisations, defined by long GRBs with measured
$z$.

\begin{table*}
\caption{Long GRBs with measured redshifts and spectral parameters not
  already in the sample of 83 bursts considered by N08.  References:
  (1) Golenetskii et al. 2008; (2) Ohno et al. 2008; (3) Barthelmy et
  al. 2008; (4) Tueller et al. 2008; (5) Golenetskii et al. 2008, GCN
  7854; (6) Golenetskii et al. 2008a; (7) Golenetskii et al. 2008b;
  (8) Meegan et al. 2008; (9) Baumgartner et al. 2008.}
\label{long} 
\begin{center}
\begin{tabular}{lllllllllll}
\hline
\noalign{\smallskip}
GRB     & z        &    $\alpha$   &  Peak flux & Range   &  $L_{\rm iso}$    &$E_{\rm peak}$&  Fluence     & range   &  $E_{\rm iso}$   &  ref   \\
        &          &               &            & (keV)   &  $10^{52}$ erg/s  &  (keV)       &  ($10^{-6}$) & keV     &  $10^{52}$ erg   &        \\
\noalign{\smallskip}
\hline
\noalign{\smallskip}
080411   & 1.03    & -1.51(0.05)   & 1.3(0.2)e-5& 20-2000 & 9.4(0.4)          &   526(63)    & 63(3.1)      & 20-2000 & 24(2)            &  1   \\
080413   & 2.433   & -1.2(0.1)     & 0.8(0.2)   & 15-1000 & 0.6(0.1)          &   584(206)   & 4.8(1.0)     & 15-1000 & 8.5(1.0)         &  2   \\
080413B  & 1.1     & -1.26(0.27)   & 18.7(0.8) &  15-150  & 1.7(0.3)          &   154(33)    & 3.2(0.1)     & 15-150  & 2.0(0.4)         &  3   \\        
080603B  & 2.69    & -1.21(0.3)    & 1.5(0.4)e-6& 20-1000 & 12(0.5)           &   376(76)    & 2.4(0.1)     & 20-1000 & 11(1.6)          &  4   \\
080605   & 1.639   & -1.03(0.07)   & 1.6(0.3)e-5& 20-2000 & 32(1.3)           &   252(19)    & 30.2(1.2)    & 20-2000 & 25.3(3.6)        &  5   \\
080607   & 3.036   & -1.08(0.07)   & 2.7(0.5)e-5& 20-4000 & 217(10)           &   1691(170)  & 89(5)        & 20-4000 & 200(13)          &  6  \\ 
080721$^{\mathrm{a}}$   & 2.591   & -0.9(0.1)     & 2.0(0.3)e-5& 20-5000 & 102(15)           &   1742(226)  & 84(6)        & 20-5000 & 120(12)          &  7   \\
080810   & 3.35    & -0.91(0.12)   & 1.9(0.2)   & 50-300  & 9.3(0.9)          &   1488(348)  & 6.9(0.5)     & 50-300  & 39(3.7)          &  8   \\
080916A  & 0.689   & -1.17(0.21)   & 2.7(0.2)   & 15-150  & 0.08(0.02)        &   161(39)    & 4.0(0.1)     & 15-150  & 1(0.2)           &  9  \\ 
\hline
\end{tabular}
\begin{list}{}{}
\item[$^{\mathrm{a}}$] Band spectrum with $\beta=-2.43\pm0.35$
\end{list}
\end{center}
\end{table*}   

\begin{table*}
\caption{Short GRBs with measured redshifts and spectral parameters.
  References: (1) Villasenor et al. 2005; (2) Golenetskii et al. 2005;
  (3) Golenetskii et al. 2006; (4) Ohno et al. 2007; (5) Golenetskii
  et al. 2007; (6) Pal'Shin et al. 2008.}
\label{short} 
\begin{center}
\begin{tabular}{lllllllllll}
\hline
\noalign{\smallskip}
GRB     & z        &    $\alpha$   &  Peak flux & Range   &  $L_{\rm iso}$    &$E_{\rm peak}$&  Fluence     & range   &  $E_{\rm iso}$   &  ref   \\
        &          &               &            & (keV)   &  $10^{52}$ erg/s  &  (keV)       &  ($10^{-6}$) & keV     &  $10^{52}$ erg   &        \\
\noalign{\smallskip}
\hline
\noalign{\smallskip}
050709  &0.16      &-0.53(0.12)    &  5.1(0.5)e-6& 2-400   &  0.05(0.01)       & 97.4(11.6)   &  0.4(0.04)   & 2-400   &   0.0033(0.0001)    &   1\\
051221  &0.5465    &-1.08(0.13)    &  4.6(1.3)e-5& 20-2000 &  6.42(0.56)       & 620(186)     &  3.2(0.9)    & 20-2000 &   0.3(0.04)      &   2\\
061006  &0.4377    &-0.62(0.2)     &  2.1e-5     & 20-2000 &  1.78(0.23)       & 955(267)     &  3.57        & 20-2000 &   0.2(0.03)      &   3\\
070714  &0.92      &-0.86(0.1)     &  2.8(0.3)   & 100-1000&  1.4(0.1)         & 2150(1113)   &  3.7         & 15-2000 &   1.1(0.1)       &   4\\
071020  &2.145     &-0.65(0.3)     &  6.0e-6     & 20-2000 &  22(1)            & 1013(205)    &  7.7         & 20-2000 &   10.2(1.5)      &   5\\ 
080913  &6.7       &-0.89(0.52)    &  1.4(0.2)   & 15-1000 &  11.4(1.5)        & 1009(200)    &  0.9         & 15-1000 &   7.14(0.9)      &   6\\
\noalign{\smallskip}
\hline
\end{tabular}
\end{center}
\end{table*}   

Amati (2006) considered two short bursts with robust redshift and peak
energy determinations that are inconsistent with the \ep--\eiso\
correlation defined by long GRBs: these two events are 3 orders of
magnitude less energetic than long GRBs of similar peak energy.  A
similar conclusion was reached by Amati (2008) for a sample of 5 short
GRBs.

However, the results presented in Sect. \ref{correl} indicate that short
and long GRBs have comparable properties in the \epo--\pf\ plane, and
that none of the 79 short events without a redshift is a clear outlier
of the \yone\ correlation: this is consistent with the hypothesis that
both populations follow the same (rest--frame) \yone\ correlation of
long GRBs.  The different region occupied by short and long
bursts in the \epo--\flu\ plane might also indicate that their 
rest--frame properties differ.

We test these possibilities for the available sample (updated to contain bursts until Sept.
2008) of short and long GRBs with measured peak energy and redshift. The long
GRB sample consists of the 83 GRBs considered by N08, which was updated by adding
the 9 long GRBs detected from March to September 2008 (Table \ref{long}).
Similarly, we searched the literature for data of all the short GRBs with measured
$z$ and spectral properties. There are a dozen GRBs with measured redshifts
that are defined in the literature as short events based on their duration.
In this class, there are few bursts composed by a short spike followed by
long--lasting (``extended'') emission (e.g., GRB~050724, GRB~061006, and
GRB~070714B, but see also Zhang et al.  2007 for the case of 060614). The
short--duration, hard, initial spike has properties similar to those of short GRBs
without the extended emission (Norris \& Gehrels 2008). Among the short GRBs
with extended emission, we consider only the two cases (GRB~061006 and
GRB~070714B) in which the short spike spectrum has a measured \epo. In
addition to these, there are 2 short GRBs, without extended emission, which
have both $z$ and \epo\ measured. These are GRB~050709 and GRB~051221.

In the sample of short GRBs (reported in Table \ref{short}) we also include the
 detected GRB~080913 ($z=6.7$), which would be classified as a long
event based on its observed duration ($\sim$8 sec), but it is intrinsically short.
This seems to be supported by the hardness of this event (Perez--Ramirez et
al. 2008, but see Greiner et al. 2008). A similar case is GRB~071020, which is
at relatively high redshift ($z=2.145$, Jakobsson et al. 2007) but has
an intrinsic duration that implies it is a member of the short class.

In Fig.\ref{ama_yone}, we show the sample of 92 long GRBs (solid filled
circles), which define the \yone\ and \ama\ correlation (left and right panels in Fig. \ref{ama_yone}, respectively).  
The solid (black) lines in Fig. \ref{ama_yone} are the best--fit functions to the correlations for the sample of long GRBs: 
\begin{equation}
E_{\rm peak}\propto L_{\rm iso}^{0.4}   (\sigma=0.27)
\end{equation}
\begin{equation}
E_{\rm peak}\propto E_{\rm iso}^{0.5}   (\sigma=0.23),
\end{equation}
where $\sigma$ is the standard deviation of the scatter of the data points perpendicular to the best--fit model relations.
Short GRBs (filled squares) are inconsistent with the \ama\ correlation (right panel in Fig. \ref{ama_yone}), while
they are consistent with the \yone\ one, defined by long GRBs (left panel in Fig.\ref{ama_yone}).  

For the 79 short and long GRBs analyzed in this paper, we do not know the redshift. However, we can test their consistency with the 
correlations of Fig. \ref{ama_yone} by assigning a redshift $z$ between 0.1 and 10. For each burst and for each redshift, we therefore compute the rest--frame peak energy $E_{\rm peak}=(1+z)E_{\rm peak}^{\rm obs}$, the bolometric isotropic energy $E_{\rm iso}=4\pi d_{L}(z)^{2} F/(1+z)$, and the bolometric isotropic luminosity $L_{\rm iso}=4\pi d_{L}(z)^{2} P$ (where $d_{L}(z)$ is the luminosity distance)
 
The curves in Fig.\ref{ama_yone} indicate where the 79 short (shaded curves) and long GRBs (solid curves) move in the 
\yone\ and \ama\ planes (left and right panels of Fig. \ref{ama_yone}, respectively) if they are assigned a redshift between 0.1 and 10. 
As a support of the tests on outliers presented in Fig. \ref{pf-ep} and Fig.~\ref{flu-ep}, data for short GRBs in
Fig. \ref{ama_yone} (left) are consistent with the \yone\ correlation
defined by long events (filled circles), while most are inconsistent with the
\ama\ correlation (right)  defined by long bursts (filled circles).

We note that in the \ama\ plane, short GRBs at $z<1.0$ are
outliers whereas the two high--redshift events GRBs (GRB~071020 at
$z=2.145$, and GRB~080913 at $z=6.7$) are consistent within the
3$\sigma$ scatter of the \ama\ correlation.

\section{Summary and conclusions}
\label{concl}

We have presented a spectral analysis of of 79 short GRBs
detected by \ba\ with peak flux $>$ 3 phot cm$^{-2}$ s$^{-1}$
(integrated in the 50-300 keV energy range). These data were compared with those of  a
representative sample of 79 long \ba\ GRBs with the same flux limit.  
For both short and long GRBs, we have analysed
the time--integrated spectra with the typical models adopted (see
e.g. Kaneko et al. 2006); a time--resolved spectral analysis has also been
 performed for long events.

Most of the short GRB spectra were reproduced more accurately by a cutoff power--law
model. For the population of long GRBs, 56\% of the spectra
are fitted by a Band and the 43\% by a cutoff power--law model. This
might reflect a genuine intrinsic difference between the spectra of
short and long GRBs or could be due to observational selection
effects. The low S/N of the high--energy part of the \ba\
spectrum does not allow us to exclude that, as for long events,
short bursts also have a high--energy spectral tail. The BGO detectors
of the GBM experiment onboard Fermi/GLAST (Meegan et al. 2008) extend
the spectral range of the NaI detectors (similar to the \ba\ LAD) to a
few tenths of MeV: this will allow us to test the nature of short GRB
high--energy emission.

The comparison of the spectral properties of short and long GRBs shows
that:
\begin{itemize}
\item the time--averaged spectrum of short GRBs is harder than that of
  long GRBs due to a harder low--energy spectral component. The peak
  energy distribution is, instead, only slightly offset towards higher energy for short than
  for long bursts. Therefore, the difference observed in the
  hardness--duration plane between the two populations is due to the
  different distribution of $\alpha$ (Fig.~\ref{histo});
\item the spectrum of short GRBs is similar to the spectrum of the
  first 1--2 secs of emission of long GRBs (Fig.~\ref{histo12sec}),
  both in terms of the low--energy spectral index and the peak
  energy. This is intriguingly consistent with the similar variability
  of short and the first few seconds of emission from long bursts (Nakar \&
  Piran 2002) and might indicate that a common mechanism operates during
  the first few seconds after the trigger for all events.
\end{itemize}

We compared the distribution of short and long GRBs in the observer
frame \epo--Peak flux and \epo--Fluence planes, where long GRBs are
known to follow well defined correlations (Ghirlanda et al. 2008, Nava
et al. 2008):
\begin{itemize}
\item short GRBs have a similar \epo\ and peak flux of long GRBs and,
  indeed, populate the same region in the \epo--Peak flux plane,
  although short bursts tend to occupy the high peak flux--peak energy
  range of the correlation defined by long GRBs
  (Fig.~\ref{pf-ep}). This suggests that short events can be
  consistent with the rest--frame \yone\ correlation defined by long
  GRBs, if their redshift distribution is similar (as also supported
  by the redshift distribution of the few short GRBs known to date).
\item short GRBs, instead, have fluences lower than those of  long events, and
  are inconsistent with the empirical correlation defined by long GRBs
  in the \epo--Fluence plane (Fig.~\ref{flu-ep}). This implies that
  the majority of short GRBs (78\%) are outliers of the \ama\
  correlation defined by long bursts.
\end{itemize}

Finally, we compared the intrinsic properties of short and long GRBs
with known redshift (Fig. \ref{ama_yone}).  Although only a few short
bursts have measured $z$ and well determined spectral properties, we
find that while short GRBs are inconsistent with the \ama\ correlation
defined by long GRBs, they are consistent with the \yone\ correlation
of the 92 GRBs with available $z$ (updated to contain data to Sept. 2008).

We conclude that the comparison of the characteristics of short GRB
with those of the first seconds of emission of long GRBs indicate that
the two population show (i) the same variability (Nakar \& Piran
2002), (ii) the same spectrum, (iii) the same luminosity, and (iv) are
consistent with the same \yone\ correlation. All of these similarities
strongly suggest a common (or similar) dissipation process -- and
possibly central engine -- acting in both classes of GRBs.  In this
respect the only difference could be the engine lifetime.

These results do not necessarily require a similar progenitor.  Both the core
collapse of a massive star and the merging of two compact objects can produce 
a black hole accreting material from a dense disc/torus.  In
this respect, one of the major differences between the two scenarios is the
absence of a supernova event accompanying short GRBs, as confirmed by the
observations. The other expected difference is in the redshift
distribution of the two populations (but see e.g. Belczynski et
al. 2008), although the detection of GRB~080913 (at $z$=6.7)
challenges this possibility.

\section*{Acknowledgments} 
We thank M. Nardini, D. Burlon and F. Tavecchio for useful discussions. We
acknowledge ASI (I/088/06/0), a 2007 PRIN--INAF grant and the MIUR for
financial support. C. Halliday is acknowledged for language editing. 
This research has made use of the data obtained through
the High Energy Astrophysics Science Archive Research Center Online Service,
provided by the NASA/Goddard Space Flight Center.

\newpage

\begin{table*}
\caption{The sample of 79 short BATSE GRBs.}
\label{tabshort} 
\begin{center}
\begin{tabular}{rlrcccll}
\hline
\noalign{\smallskip}
Trig. &~~~~~~T$_{90}$& P~~~~~~~~          & $\alpha$          & $E_0$                  & $\chi^2$(dof)   &  Fluence      &  Peak flux   \\
      &~~~~~~s       & phot/(cm$^2$ s)    &                   & keV                    &                 &  erg/cm$^2$   & erg/(cm$^2$s)\\
\noalign{\smallskip}
\hline
\noalign{\smallskip}
6293  &  0.192$\pm$0.091 & 88.53$\pm$1.00 &  -1.27$\pm$0.02   &                        &    1.216(109)   &    4.56E-6   & $>$5.74E-5  \\
298   &  0.455$\pm$0.065 & 56.13$\pm$1.27 &  -0.57$\pm$0.92   &     85.38$\pm$ 64.90   &    1.113(102)   &    1.99E-7   &    1.43E-5  \\
3412  &  0.068$\pm$0.006 & 54.82$\pm$0.76 &  -1.31$\pm$0.52   &    110.20$\pm$ 80.98   &    0.892(103)   &    2.62E-7   &    1.91E-5  \\
6668  &  0.116$\pm$0.006 & 39.12$\pm$0.61 &  -0.39$\pm$0.49   &    126.80$\pm$ 62.57   &    1.184(107)   &    4.99E-7   &    1.18E-5  \\
444   &  0.256$\pm$0.091 & 28.55$\pm$0.76 &  -0.87$\pm$0.23   &    113.50$\pm$ 28.39   &    1.132(102)   &    5.07E-7   &    8.04E-6  \\
2514  &  0.200$\pm$0.094 & 28.40$\pm$0.74 &  -0.81$\pm$0.14   &    163.30$\pm$ 25.95   &    1.129(100)   &    1.12E-6   &    8.99E-6  \\
3152  &  1.793$\pm$0.066 & 25.34$\pm$0.72 &  -0.40$\pm$0.09   &    683.70$\pm$116.50   &    1.175(107)   &    6.55E-6   &    4.64E-5  \\
5561  &  0.104$\pm$0.011 & 19.28$\pm$0.45 &  -1.20$\pm$1.48   &     48.51$\pm$ 25.00   &    0.956(108)   &    1.65E-7   &    8.69E-6  \\
3087  &  1.152$\pm$0.091 & 18.68$\pm$0.58 &  -1.19$\pm$0.15   &    273.10$\pm$ 74.50   &    1.103( 76)   &    2.89E-6   &    7.02E-6  \\
2273  &  0.224$\pm$0.066 & 18.59$\pm$0.55 &  -0.18$\pm$0.45   &    132.70$\pm$ 49.46   &    0.886(100)   &    3.88E-7   &    6.26E-6  \\
7281  &  1.664$\pm$0.143 & 16.83$\pm$0.42 &  -0.83$\pm$0.15   &    123.30$\pm$ 18.60   &    1.296(107)   &    2.21E-6   &    4.80E-6  \\
2068  &  0.591$\pm$0.060 & 15.63$\pm$0.59 &  -0.22$\pm$0.26   &     97.07$\pm$ 22.85   &    1.210(107)   &    3.91E-7   &    4.19E-6  \\
2125  &  0.223$\pm$0.013 & 15.42$\pm$0.56 &  -0.48$\pm$0.30   &    240.50$\pm$ 90.00   &    0.844(102)   &    4.57E-7   &    7.43E-6  \\
3173  &  0.208$\pm$0.025 & 14.90$\pm$0.58 &  -1.00$\pm$0.18   &    559.60$\pm$281.65   &    1.356(105)   &    6.69E-7   &    9.52E-6  \\
2679  &  0.256$\pm$0.091 & 13.73$\pm$0.51 &  -0.32$\pm$0.13   &    650.20$\pm$149.25   &    1.363(107)   &    3.14E-6   &    2.72E-5  \\
1553  &  0.960$\pm$0.143 & 13.70$\pm$0.52 &  -0.87$\pm$0.11   &    764.00$\pm$183.60   &    1.173( 96)   &    6.62E-6   &    1.35E-5  \\
6123  &  0.186$\pm$0.042 & 12.83$\pm$0.42 &  -0.23$\pm$1.64   &     76.66$\pm$ 49.00   &    1.107(108)   &    1.11E-7   &    3.10E-6  \\
6635  &  1.152$\pm$0.143 & 12.05$\pm$0.39 &  -1.74$\pm$0.15   &    129.50$\pm$ 32.70   &    1.014( 91)   &    2.76E-6   &    6.57E-6  \\
1088  &  0.192$\pm$0.091 & 11.92$\pm$0.55 &   0.10$\pm$2.11   &     68.08$\pm$ 61.79   &    1.186(104)   &    7.41E-8   &    2.80E-6  \\
1453  &  0.192$\pm$0.453 & 11.89$\pm$0.51 &  -0.16$\pm$0.65   &     94.20$\pm$ 48.00   &    0.812(108)   &    1.80E-7   &    3.17E-6  \\
6535  &  1.664$\pm$0.143 & 11.88$\pm$0.38 &  -0.97$\pm$0.08   &   1175.60$\pm$384.27   &    1.391(108)   &    7.36E-6   &    1.47E-5  \\
2320  &  0.608$\pm$0.041 & 11.03$\pm$0.47 &  -0.58$\pm$0.19   &    129.00$\pm$ 26.10   &    0.794(103)   &    7.57E-7   &    3.23E-6  \\
2933  &  0.320$\pm$0.091 & 10.77$\pm$0.44 &   0.22$\pm$0.62   &    130.20$\pm$ 55.94   &    1.429(107)   &    3.42E-6   &    4.33E-6  \\
7939  &  1.039$\pm$0.072 & 10.77$\pm$0.38 &  -0.41$\pm$0.15   &     99.73$\pm$ 12.96   &    1.193( 82)   &    2.53E-6   &    2.86E-6  \\
2614  &  0.296$\pm$0.057 & 10.49$\pm$0.52 &  -1.00$\pm$0.18   &    469.60$\pm$222.80   &    0.836(108)   &    6.08E-7   &    5.84E-6  \\
2715  &  0.384$\pm$0.091 & 10.47$\pm$0.50 &   0.08$\pm$0.11   &    562.80$\pm$ 85.20   &    1.049(108)   &    7.69E-6   &    3.30E-5  \\
2896  &  0.456$\pm$0.033 & 10.44$\pm$0.48 &  -0.87$\pm$0.26   &     79.94$\pm$ 18.19   &    1.072(106)   &    7.53E-7   &    2.89E-6  \\
7784  &  1.918$\pm$1.995 & 10.29$\pm$0.34 &  -0.83$\pm$0.35   &    140.20$\pm$ 54.30   &    1.432(108)   &    5.63E-7   &    3.05E-6  \\
2317  &  0.896$\pm$0.091 & 9.73$\pm$0.46  &  -0.53$\pm$0.25   &     73.46$\pm$ 13.12   &    1.249( 65)   &    1.04E-6   &    2.41E-6  \\
2834  &  0.680$\pm$0.011 & 8.79$\pm$0.44  &  -0.54$\pm$0.24   &    407.60$\pm$168.80   &    1.165( 85)   &    1.36E-6   &    6.90E-6  \\
6679  &  1.408$\pm$0.091 & 8.62$\pm$0.35  &  -0.61$\pm$0.27   &    318.90$\pm$141.60   &    1.409(107)   &    9.39E-7   &    4.91E-6  \\
6527  &  1.856$\pm$0.516 & 8.47$\pm$0.38  &  -1.32$\pm$0.21   &     80.36$\pm$ 15.60   &    1.090( 95)   &    3.33E-6   &    3.25E-6  \\
7353  &  0.249$\pm$0.004 & 8.47$\pm$0.38  &   0.00$\pm$0.22   &    615.80$\pm$197.40   &    1.181(107)   &    4.19E-6   &    2.72E-5  \\
5277  &  0.496$\pm$0.023 & 8.14$\pm$0.33  &   0.29$\pm$0.24   &    208.40$\pm$ 30.81   &    0.885(106)   &    1.54E-6   &    6.46E-6  \\
8104  &  0.384$\pm$0.091 & 8.13$\pm$0.30  &   0.42$\pm$1.35   &    110.60$\pm$ 70.37   &    0.774(107)   &    2.20E-7   &    3.04E-6  \\
2330  &  0.804$\pm$0.009 & 8.03$\pm$0.39  &  -0.86$\pm$0.29   &    616.90$\pm$491.30   &    0.961( 75)   &    1.02E-6   &    6.54E-6  \\
6263  &  1.984$\pm$0.181 & 7.99$\pm$0.31  &  -0.36$\pm$0.64   &     69.14$\pm$ 30.59   &    1.054(107)   &    3.78E-7   &    1.91E-6  \\
5339  &  0.832$\pm$0.091 & 7.77$\pm$0.33  &  -0.40$\pm$0.10   &    567.90$\pm$ 99.64   &    0.732( 93)   &    4.95E-6   &    1.12E-5  \\
603   &  1.472$\pm$0.272 & 7.50$\pm$0.56  &  -0.71$\pm$0.63   &    155.30$\pm$ 93.62   &    1.004( 85)   &    3.78E-7   &    2.36E-6  \\
6368  &  0.896$\pm$0.326 & 7.24$\pm$0.34  &  -1.37$\pm$0.18   &                        &    0.997(108)   &    3.21E-7   & $>$4.26E-6  \\
6606  &  0.704$\pm$0.389 & 7.16$\pm$0.29  &  -1.77$\pm$0.20   &                        &    0.973(108)   &    5.02E-7   & $>$3.04E-6  \\
3642  &  0.704$\pm$0.091 & 6.83$\pm$0.31  &   0.21$\pm$0.88   &     89.97$\pm$ 58.42   &    1.262(107)   &    2.92E-7   &    1.93E-6  \\
6671  &  0.256$\pm$0.091 & 6.71$\pm$0.31  &  -1.39$\pm$0.13   &                        &    0.937(100)   &    5.36E-7   & $>$3.84E-6  \\
5647  &  1.088$\pm$0.326 & 6.50$\pm$0.32  &  -0.06$\pm$0.80   &    108.50$\pm$115.16   &    1.366(107)   &    1.74E-7   &    1.95E-6  \\
7375  &  0.311$\pm$0.073 & 6.40$\pm$0.31  &  -0.47$\pm$0.87   &    267.90$\pm$200.05   &    1.039(101)   &    3.19E-7   &    3.46E-6  \\
677   &  0.055$\pm$0.008 & 6.21$\pm$0.44  &   0.65$\pm$1.29   &    127.20$\pm$168.26   &    0.751(105)   &    1.22E-7   &    3.18E-6  \\
1076  &  0.161$\pm$0.016 & 6.18$\pm$0.44  &  -2.46$\pm$0.33   &                        &    1.417( 89)   &    1.20E-7   & $>$2.16E-6  \\
936   &  1.438$\pm$0.065 & 5.85$\pm$0.44  &  -0.84$\pm$0.26   &    341.50$\pm$179.45   &    1.069(104)   &    7.03E-7   &    2.91E-6  \\
5607  &  1.088$\pm$0.091 & 5.85$\pm$0.30  &  -0.71$\pm$0.23   &    426.20$\pm$199.45   &    1.150( 82)   &    1.19E-6   &    3.97E-6  \\
7142  &  0.969$\pm$0.064 & 5.81$\pm$0.28  &   0.94$\pm$0.33   &    124.10$\pm$ 12.79   &    0.953(107)   &    1.42E-6   &    3.50E-6  \\
4955  &  0.464$\pm$0.036 & 5.73$\pm$0.31  &  -1.04$\pm$0.45   &    298.20$\pm$371.80   &    1.176(107)   &    2.71E-7   &    2.33E-6  \\
4776  &  0.448$\pm$0.091 & 5.54$\pm$0.28  &  -0.19$\pm$0.32   &    232.70$\pm$ 88.45   &    1.152(107)   &    6.90E-8   &    3.27E-6  \\
7813  &  0.564$\pm$0.164 & 5.37$\pm$0.29  &  -2.68$\pm$0.17   &                        &    1.053(108)   &    5.59E-7   & $>$1.94E-6  \\
1760  &  0.576$\pm$0.143 & 5.27$\pm$0.35  &  -0.25$\pm$0.28   &    188.70$\pm$ 56.95   &    1.027(105)   &    6.18E-7   &    2.37E-6  \\
7378  &  1.247$\pm$0.077 & 5.25$\pm$0.33  &  -0.52$\pm$0.16   &    536.20$\pm$153.35   &    1.465(107)   &    2.60E-6   &    5.87E-6  \\
4660  &  1.168$\pm$0.080 & 5.15$\pm$0.29  &   0.56$\pm$0.21   &    161.70$\pm$ 23.80   &    0.919( 87)   &    1.92E-6   &    3.53E-6  \\
5533  &  0.768$\pm$0.091 & 5.12$\pm$0.30  &   0.02$\pm$0.15   &    335.20$\pm$ 60.15   &    0.971( 87)   &    2.91E-7   &    6.26E-6  \\
7078  &  0.448$\pm$0.091 & 5.11$\pm$0.42  &  -3.60$\pm$0.45   &                        &    0.920(108)   &    1.73E-7   & $>$2.90E-6  \\
\noalign{\smallskip}
\hline
\end{tabular}
\end{center}
\end{table*}

\setcounter{table}{5}
\begin{table*}
\caption{continue....}
\begin{center}
\begin{tabular}{rlrcccll}
\hline
\noalign{\smallskip}
Trig. &~~~~~~T$_{90}$& P~~~~~~~~          & $\alpha$          & $E_0$                  & $\chi^2$(dof)   &  Fluence      &  Peak flux   \\
      &~~~~~~s       &~phot/(cm$^2$ s)    &                   & keV                    &                 &  erg/cm$^2$   & erg/(cm$^2$s)\\
\noalign{\smallskip}
\hline
\noalign{\smallskip}
5527  &  0.820$\pm$0.008 & 5.04$\pm$0.26  &  -0.34$\pm$0.11   &    489.30$\pm$ 88.30   &    0.760( 90)   &    3.73E-6   &    6.41E-6  \\
3735  &  1.301$\pm$0.091 & 4.83$\pm$0.29  &   0.00$\pm$0.18   &    301.70$\pm$ 55.05   &    1.286(107)   &    2.60E-6   &    4.91E-6  \\
3297  &  0.272$\pm$0.023 & 4.45$\pm$0.33  &  -0.83$\pm$0.37   &    496.80$\pm$501.70   &    1.198(106)   &    4.90E-7   &    3.07E-6  \\
2952  &  0.680$\pm$0.018 & 4.37$\pm$0.34  &  -0.69$\pm$0.25   &    570.20$\pm$312.15   &    0.791(107)   &    8.76E-7   &    4.13E-6  \\
5599  &  0.598$\pm$0.043 & 4.24$\pm$0.26  &  -0.79$\pm$0.30   &    664.70$\pm$637.40   &    1.234(106)   &    8.25E-7   &    4.07E-6  \\
5529  &  1.015$\pm$0.129 & 4.23$\pm$0.29  &   1.37$\pm$0.96   &     65.65$\pm$ 22.09   &    1.015(106)   &    2.95E-7   &    1.31E-6  \\
7133  &  1.079$\pm$0.37  & 4.08$\pm$0.26  &  -0.14$\pm$0.29   &    135.80$\pm$ 36.25   &    1.115(107)   &    6.01E-7   &    1.43E-6  \\
7793  &  1.093$\pm$0.04  & 3.99$\pm$0.27  &  -0.05$\pm$0.22   &    470.90$\pm$126.35   &    1.054(106)   &    4.34E-6   &    7.56E-6  \\
2377  &  0.496$\pm$0.011 & 3.98$\pm$0.33  &   0.06$\pm$0.26   &    229.30$\pm$ 55.10   &    0.875(100)   &    6.90E-7   &    2.91E-6  \\
3606  &  1.824$\pm$0.066 & 3.95$\pm$0.26  &   0.19$\pm$0.35   &    175.90$\pm$ 49.60   &    1.216(102)   &    1.72E-6   &    2.26E-6  \\
3113  &  0.976$\pm$0.023 & 3.90$\pm$0.35  &  -0.78$\pm$0.16   &    690.00$\pm$316.25   &    1.145( 90)   &    1.54E-6   &    3.95E-6  \\
6715  &  0.452$\pm$0.027 & 3.71$\pm$0.26  &  -0.25$\pm$0.78   &    206.20$\pm$187.77   &    1.178(107)   &    4.34E-7   &    1.83E-6  \\
575   &  0.413$\pm$0.022 & 3.70$\pm$0.46  &   0.17$\pm$0.87   &    121.40$\pm$ 63.56   &    0.890(106)   &    1.71E-7   &    1.35E-6  \\
2217  &  0.656$\pm$0.029 & 3.56$\pm$0.31  &   0.36$\pm$0.27   &    281.00$\pm$ 93.35   &    1.234( 73)   &    1.46E-6   &    4.97E-6  \\
3921  &  0.464$\pm$0.161 & 3.52$\pm$0.24  &   0.36$\pm$0.48   &    179.90$\pm$ 66.60   &    1.086(106)   &    5.42E-7   &    2.39E-6  \\
5206  &  0.304$\pm$0.023 & 3.46$\pm$0.28  &  -1.23$\pm$0.09   &                        &    1.219(107)   &    3.81E-7   & $>$2.34E-6  \\
2918  &  0.448$\pm$0.091 & 3.44$\pm$0.34  &  -0.60$\pm$0.63   &    252.50$\pm$195.90   &    1.085(100)   &    1.77E-7   &    1.59E-6  \\
3940  &  0.576$\pm$0.091 & 3.19$\pm$0.22  &  -0.33$\pm$0.44   &    101.80$\pm$ 40.67   &    1.187( 97)   &    2.50E-7   &    8.64E-7  \\
7912  &  1.856$\pm$0.707 & 3.10$\pm$0.25  &  -0.28$\pm$0.26   &    150.90$\pm$ 47.65   &    1.236(107)   &    8.05E-7   &    1.11E-6  \\
6341  &  1.920$\pm$0.707 & 3.05$\pm$0.28  &  -0.25$\pm$0.29   &    332.00$\pm$143.20   &    0.878(107)   &    1.34E-6   &    2.64E-6  \\
3359  &  0.344$\pm$0.025 & 3.01$\pm$0.25  &   0.67$\pm$0.90   &    121.00$\pm$ 74.79   &    1.037(104)   &    2.35E-7   &    1.46E-6  \\
\noalign{\smallskip}
\hline
\end{tabular}
\end{center}
\end{table*}

\newpage
\begin{table*}
\caption{The sample of 79 long BATSE GRBs.}
\label{tablong}
\begin{center}
\begin{tabular}{rccrrrlll}
\hline
\noalign{\smallskip}
Trig.  & T$_{90}$ & P               & $\alpha$   & $\beta$           &    $E_0$               & $\chi^2$(dof)&  Fluence     & Peak flux \\
       &    s     & phot/(cm$^2$ s) &            &                   &   keV                  &              &  erg/cm$^2$  &erg/(cm$^2$s)\\
\noalign{\smallskip}
\hline
\noalign{\smallskip}
 160  &   17.024  & 4.21     &   -0.44$\pm$0.07  &                   &    112.16$\pm$7.16     &   1.006(69)  & 5.47E-6     & 1.1707E-6\\
 543  &   4.8640  & 11.158   &   -0.87$\pm$0.05  & -2.42$\pm$0.10    &    219.22$\pm$18.01    &   0.870(82)  & 1.27E-5     & 5.7328E-6\\
 907  &   158.08  & 3.744    &    0.07$\pm$0.09  & -2.87$\pm$0.16    &     92.76$\pm$6.49     &   1.245(68)  & 7.07E-6     & 1.3548E-6\\
 973  &   89.984  & 5.707    &   -1.03$\pm$0.04  & -2.15$\pm$0.04    &    278.59$\pm$22.03    &   1.132(98)  & 4.59E-5     & 3.7020E-6\\
1122  &   18.752  & 13.787   &   -0.91$\pm$0.04  & -2.54$\pm$0.05    &    149.82$\pm$7.85     &   0.955(85)  & 3.05E-5     & 5.6131E-6\\
1157  &   170.56  & 12.187   &   -1.01$\pm$0.05  & -2.23$\pm$0.06    &    206.93$\pm$19.32    &   0.855(90)  & 2.61E-5     & 6.6804E-6\\
1159  &   18.240  & 3.727    &   -0.81$\pm$0.09  &                   &    257.21$\pm$36.81    &   0.708(58)  & 1.84E-6     & 1.5362E-6\\
1425  &   10.432  & 9.52     &   -1.52$\pm$0.03  &                   &    346.21$\pm$26.32    &   0.992(76)  & 1.20E-5     & 4.4592E-6\\
1625  &   16.128  & 28.061   &   -0.82$\pm$0.01  & -2.43$\pm$0.06    &    393.90$\pm$14.21    &   1.202(103) & 1.00E-4     & 2.1800E-5\\
1886  &   275.71  & 16.683   &   -0.43$\pm$0.02  & -2.25$\pm$0.05    &    315.04$\pm$11.47    &   1.208(103) & 7.97E-5     & 1.8097E-5\\
1922  &   16.192  & 3.532    &   -1.03$\pm$0.13  &                   &    168.01$\pm$31.15    &   0.691(48)  & 1.44E-6     & 1.1215E-6\\
2037  &   6.2720  & 8.355    &   -1.03$\pm$0.05  &                   &    709.27$\pm$117.6    &   0.924(78)  & 6.75E-6     & 6.1974E-6\\
2067  &   30.848  & 18.919   &   -0.52$\pm$0.01  & -3.15$\pm$0.07    &    173.18$\pm$2.77     &   1.177(97)  & 7.44E-5     & 7.7100E-6\\
2083  &   15.168  & 46.554   &   -1.17$\pm$0.02  & -2.41$\pm$0.03    &    238.88$\pm$9.69     &   3.283(96)  & 7.47E-5     & 2.2391E-5\\
2367  &   18.816  & 4.6      &    0.45$\pm$0.18  &                   &     92.73$\pm$13.21    &   0.762(36)  & 5.83E-7     & 1.1793E-6\\
2393  &   5.1340  & 4.388    &   -1.29$\pm$0.27  & -2.86$\pm$0.06    &     61.29$\pm$16.24    &   0.841(53)  & 8.16E-6     & 2.0287E-6\\
2446  &   8.2560  & 4.378    &   -0.64$\pm$0.05  &                   &    207.11$\pm$14.64    &   1.067(73)  & 6.65E-6     & 1.6800E-6\\
2537  &   4.8000  & 27.283   &   -1.38$\pm$0.04  & -2.89$\pm$0.07    &    154.85$\pm$10.03    &   1.078(75)  & 2.69E-5     & 1.0752E-5\\
2611  &   12.212  & 35.05    &   -1.07$\pm$0.02  &                   &    662.87$\pm$46.27    &   1.666(88)  & 1.53E-5     & 2.3439E-5\\
2793  &   6.9760  & 5.086    &   -0.53$\pm$0.05  &                   &    509.84$\pm$46.32    &   0.820(86)  & 8.19E-6     & 5.2682E-6\\
2890  &   51.584  & 3.008    &   -0.98$\pm$0.04  &                   &    988.18$\pm$118.43   &   1.033(103) & 3.00E-5     & 3.1272E-6\\
2913  &   22.912  & 5.738    &   -1.31$\pm$0.14  &                   &    168.67$\pm$36.05    &   0.961(56)  & 4.95E-6     & 2.0695E-6\\
2958  &   36.896  & 3.939    &   -0.95$\pm$0.11  &                   &    133.66$\pm$15.29    &   0.948(53)  & 3.66E-6     & 1.1608E-6\\
2993  &   44.800  & 4.255    &   -1.00$\pm$0.03  &                   &   2065.78$\pm$298.74   &   0.938(102) & 4.02E-5     & 8.2881E-6\\
2994  &   48.576  & 15.349   &   -1.03$\pm$0.01  &                   &   1374.62$\pm$93.16    &   0.958(100) & 8.01E-5     & 1.9596E-5\\
3039  &   3.6350  & 9.048    &   -0.70$\pm$0.06  &                   &    102.33$\pm$5.49     &   1.298(61)  & 4.77E-6     & 2.4114E-6\\
3110  &   10.176  & 4.449    &   -0.05$\pm$0.06  &                   &    285.55$\pm$19.04    &   0.962(90)  & 1.23E-5     & 3.9332E-6\\
3138  &   5.1840  & 16.833   &   -1.32$\pm$0.03  &                   &    277.05$\pm$17.82    &   1.515(75)  & 1.30E-5     & 6.6159E-6\\
3178  &   39.936  & 14.583   &   -1.09$\pm$0.01  &                   &    782.28$\pm$35.70    &   1.638(103) & 6.13E-5     & 1.0825E-5\\
3255  &   34.880  & 12.667   &   -1.89$\pm$0.04  &                   &    325.08$\pm$39.89    &   1.334(77)  & 3.42E-5     & 1.2624E-5\\
3269  &   13.888  & 8.365    &   -0.66$\pm$0.04  &                   &    516.69$\pm$48.49    &   1.091(89)  & 1.05E-5     & 7.3692E-6\\
3287  &   33.408  & 7.714    &   -1.19$\pm$0.11  &                   &    281.28$\pm$56.33    &   0.924(74)  & 1.43E-5     & 2.9709E-6\\
3306  &   108.51  & 3.496    &   -1.13$\pm$0.41  & -2.28$\pm$0.12    &     96.75$\pm$56.07    &   0.617(78)  & 2.63E-5     & 1.6435E-6\\
3352  &   46.336  & 3.84     &   -0.81$\pm$0.05  & -2.84$\pm$0.15    &    129.93$\pm$8.327    &   0.965(75)  & 2.74E-5     & 1.3351E-6\\
3436  &   40.000  & 3.89     &   -1.09$\pm$0.09  & -2.26$\pm$0.17    &    227.08$\pm$43.66    &   0.875(80)  & 1.37E-5     & 2.0817E-6\\
3648  &   57.088  & 5.907    &   -1.08$\pm$0.11  & -2.65$\pm$0.13    &    126.45$\pm$18.19    &   0.955(67)  & 1.67E-5     & 2.2141E-6\\
3776  &   11.072  & 5.897    &   -0.46$\pm$0.07  &                   &    113.65$\pm$7.41     &   0.927(64)  & 6.40E-6     & 1.6459E-6\\
3905  &   24.256  & 4.675    &   -1.21$\pm$0.06  &                   &    352.68$\pm$53.12    &   1.313(79)  & 1.08E-5     & 1.9777E-6\\
4048  &   13.696  & 4.864    &   -0.51$\pm$0.08  & -2.41$\pm$0.15    &    225.22$\pm$26.67    &   0.843(83)  & 1.38E-5     & 3.2398E-6\\
4350  &   52.000  & 3.521    &   -1.92$\pm$0.07  &                   &    642.17$\pm$227.12   &   0.957(76)  & 1.77E-5     & 3.5523E-6\\
4710  &   9.9840  & 3.012    &   -0.25$\pm$0.95  & -2.12$\pm$0.08    &     51.53$\pm$36.96    &   1.036(57)  & 4.34E-6     & 1.5089E-6\\
5526  &   72.448  & 3.779    &   -1.34$\pm$0.05  &                   &    490.21$\pm$93.28    &   1.168(82)  & 1.98E-5     & 1.8081E-6\\
5530  &   4.9550  & 6.695    &   -1.19$\pm$0.11  &                   &     82.07$\pm$7.92     &   1.260(46)  & 5.14E-6     & 2.3486E-6\\
5563  &   4.8900  & 22.704   &   -1.01$\pm$0.08  & -2.41$\pm$0.09    &    175.29$\pm$23.56    &   1.191(72)  & 8.38E-6     & 1.0300E-5\\
5601  &   19.456  & 4.9375   &   -0.56$\pm$0.09  & -2.52$\pm$0.18    &    157.04$\pm$18.81    &   1.207(77)  & 1.32E-5     & 2.3136E-6\\
5628  &   15.872  & 8.9689   &   -1.35$\pm$0.03  &                   &    384.96$\pm$33.81    &   0.975(78)  & 1.42E-5     & 3.9267E-6\\
5704  &   10.048  & 43.927   &   -1.52$\pm$0.06  &                   &    320.81$\pm$56.10    &   0.741(73)  & 1.54E-5     & 2.0545E-5\\
5711  &   2.2400  & 41.245   &   -1.04$\pm$0.03  & -2.04$\pm$0.09    &    599.11$\pm$69.74    &   1.069(95)  & 2.16E-5     & 4.2236E-5\\
5773  &   31.488  & 15.209   &   -0.28$\pm$0.02  &                   &    103.86$\pm$1.49     &   2.148(80)  & 4.38E-5     & 4.1832E-6\\
5955  &   11.648  & 3.5974   &   -0.76$\pm$0.22  &                   &     95.97$\pm$19.25    &   0.921(38)  & 1.17E-6     & 9.5567E-7\\
\noalign{\smallskip}
\hline
\end{tabular}
\end{center}
\end{table*}

\setcounter{table}{6}
\begin{table*}
\caption{continue....}
\begin{center}
\begin{tabular}{rccrrrlll}
\hline
\noalign{\smallskip}
Trig.  & T$_{90}$ & P               & $\alpha$   & $\beta$           &    $E_0$               & $\chi^2$(dof)&  Fluence    & Peak flux \\
       &    s     & phot/(cm$^2$ s) &            &                   &   keV                  &              &  erg/cm$^2$ &erg/(cm$^2$s)\\
\noalign{\smallskip}
\hline
\noalign{\smallskip}
6100  &   16.256  & 19.4123  &   -0.98$\pm$0.02  & -2.27$\pm$0.08    &    491.21$\pm$27.50    &   1.228(107) & 7.12E-5     & 1.6291E-5\\
6235  &   4.0320  & 21.7417  &   -0.74$\pm$0.03  &                   &    294.73$\pm$14.46    &   1.092(90)  & 1.71E-5     & 1.0389E-5\\
6251  &   3.0080  & 7.5941   &   -1.05$\pm$0.10  &                   &    497.31$\pm$128.32   &   1.018(61)  & 3.16E-6     & 4.2470E-6\\
6266  &   37.568  & 4.0372   &   -0.61$\pm$0.04  & -2.63$\pm$0.20    &    229.27$\pm$16.52    &   1.035(90)  & 3.00E-5     & 2.2454E-6\\
6336  &   6.5280  & 15.6576  &   -1.08$\pm$0.02  &                   &   1214.97$\pm$88.84    &   0.964(101) & 3.91E-5     & 1.6621E-5\\
6400  &   14.080  & 4.0832   &   -1.72$\pm$0.05  &                   &                        &   0.907(53)  & 2.15E-5     &$>$1.7827E-6\\
6422  &   6.5920  & 8.2515   &   -1.27$\pm$0.11  & -3.44$\pm$0.14    &     66.92$\pm$6.48     &   1.944(55)  & 9.46E-6     & 3.3758E-6\\
6546  &   6.7840  & 3.3192   &   -0.45$\pm$0.29  &                   &     47.65$\pm$7.69     &   0.500(32)  & 1.17E-6     & 8.2697E-7\\
6560  &   36.800  & 11.8054  &   -0.67$\pm$0.04  &                   &    143.28$\pm$6.96     &   1.279(73)  & 1.61E-5     & 3.5545E-6\\
6593  &   31.232  & 10.1204  &   -0.93$\pm$0.03  & -2.28$\pm$0.07    &    225.64$\pm$15.44    &   1.236(96)  & 5.32E-5     & 5.6690E-6\\
6814  &   19.264  & 3.5187   &   -0.31$\pm$0.38  & -2.15$\pm$0.09    &     80.42$\pm$26.71    &   1.002(59)  & 5.76E-6     & 1.8153E-6\\
6816  &   35.136  & 5.5426   &   -1.68$\pm$0.09  &                   &    499.27$\pm$166.46   &   1.093(63)  & 1.62E-5    & 3.3415E-6\\
6930  &   36.800  & 5.7627   &   -0.62$\pm$0.25  & -2.34$\pm$0.08    &     65.41$\pm$15.49    &   0.924(66)  & 1.09E-5    & 2.3424E-6\\
7185  &   147.45  & 5.8478   &   -1.66$\pm$0.12  &                   &    153.26$\pm$30.25    &   1.295(40)  & 3.48E-5    & 3.7113E-6\\
7255  &   10.304  & 5.391    &   -0.61$\pm$0.05  & -2.73$\pm$0.18    &    174.32$\pm$13.16    &   1.117(82)  & 1.42E-5    & 2.3779E-6\\
7318  &   14.976  & 4.0609   &   -0.58$\pm$0.03  &                   &    469.06$\pm$30.66    &   1.053(93)  & 2.05E-5    & 3.5686E-6\\
7329  &   3.1360  & 4.2126   &    0.36$\pm$0.09  & -2.59$\pm$0.24    &    181.90$\pm$16.54    &   0.964(81)  & 8.29E-6    & 4.3176E-6\\
7374  &   12.160  & 4.6678   &    0.32$\pm$0.10  &                   &     77.58$\pm$4.71     &   0.626(61)  & 3.73E-6    & 1.2204E-6\\
7429  &   17.856  & 3.6964   &    0.22$\pm$0.07  &                   &    100.66$\pm$4.95     &   1.057(70)  & 1.01E-5    & 1.1517E-6\\
7464  &   41.088  & 5.797    &   -0.78$\pm$0.03  & -2.73$\pm$0.22    &    342.83$\pm$19.63    &   1.188(97)  & 4.26E-5    & 3.6946E-6\\
7515  &   328.44  & 3.2372   &   -0.47$\pm$0.10  & -2.28$\pm$0.11    &    130.60$\pm$16.54    &   1.111(78)  & 1.69E-5    & 1.7024E-6\\
7518  &   12.736  & 3.7771   &    0.26$\pm$0.18  &                   &    142.20$\pm$19.32    &   0.670(74)  & 5.84E-6    & 1.7344E-6\\
7678  &   42.752  & 11.7202  &   -0.81$\pm$0.03  & -2.53$\pm$0.09    &    290.72$\pm$14.39    &   1.473(107) & 1.04E-4    & 7.0047E-6\\
7906  &   15.168  & 91.4818  &   -1.09$\pm$0.01  & -2.27$\pm$0.02    &    420.65$\pm$11.75    &   2.032(107) & 3.19E-4    & 6.3174E-5\\
7932  &   67.392  & 3.1139   &   -1.23$\pm$0.08  &                   &    181.15$\pm$21.32    &   1.712(78)  & 1.90E-5    & 1.0716E-6\\
7954  &   15.040  & 56.9544  &   -0.93$\pm$0.02  & -2.80$\pm$0.12    &    212.67$\pm$8.64     &   1.529(95)  & 4.18E-5    & 2.3804E-5\\
7998  &   10.240  & 4.505    &   -0.53$\pm$0.19  &                   &     57.98$\pm$7.61     &   1.016(36)  & 2.32E-6    & 1.1128E-6\\
8008  &   22.656  & 9.0618   &   -0.46$\pm$0.03  & -2.42$\pm$0.09    &    293.21$\pm$14.55    &   1.285(97)  & 6.10E-5    & 7.9296E-6\\
8099  &   15.488  & 8.5515   &   -1.59$\pm$0.07  &                   &    181.37$\pm$21.05    &   1.460(55)  & 8.25E-6    & 4.4529E-6\\
\noalign{\smallskip}
\hline
\end{tabular}
\end{center}
\end{table*}   

\end{document}